\documentclass[a4paper,fleqn]{cas-sc}

\usepackage[authoryear,longnamesfirst]{natbib}
\usepackage[utf8]{inputenc}
\usepackage{amsmath,amssymb,physics,cancel,graphicx,subcaption,listings,matlab-prettifier}
\usepackage{nccmath}
\usepackage[english]{babel}
\usepackage{amsthm}
\graphicspath{ {./images/} }
\usepackage{xcolor}
\usepackage{placeins}
\usepackage{enumitem}
\usepackage{soul, hanging}
\usepackage{rotating}
\usepackage{natbib}
\usepackage{bbding}
\def\tsc#1{\csdef{#1}{\textsc{\lowercase{#1}}\xspace}}
\tsc{WGM}
\tsc{QE}

\begin{document}
\let\WriteBookmarks\relax
\def\floatpagepagefraction{1}
\def\textpagefraction{.001}

\shorttitle{Household coping mechanisms under grid failure: Evidence from a high electrification context in Lebanon}    

\shortauthors{Olleik et al.}  

\title [mode = title]{Household coping mechanisms under grid failure: Evidence from a high electrification context in Lebanon}  



%

\author[1]{Majd Olleik}[orcid=0009-0006-4082-4102]

\cormark[1]


\ead{mo10@aub.edu.lb}



\affiliation[1]{organization={Department of Industrial Engineering and Management, American University of Beirut},
            city={Beirut},
            country={Lebanon}}

\author[2,3]{Haytham M. Dbouk}





\affiliation[2]{organization={Department of Electrical and Communication Engineering, Phoenicia University, Lebanon},
                addressline={District of Zahrani}, 
                city={South Lebanon},
                country={Lebanon}}
                
\affiliation[3]{organization={Innovating Green Technology},
            city={Beirut},
            country={Lebanon}}



\author[4]{Anne Neumann}[]
\affiliation[4]{organization={Department of Industrial Economics and Technology Management, The Norwegian University of Science and Technology},
            city={Trondheim},
            country={Norway}}

\author[1]{Elsa {Bou Gebrael}}
\author[5, 4]{Sebastian Zwickl-Bernhard} [orcid=0000-0002-8599-6278]
\affiliation[5]{organization={Energy Economics Group (EEG), Technische Universität Wien},
            city={Wien},
            country={Austria}}


\begin{abstract}
Despite near-universal electrification in many countries, electricity supply shortages continue to shape household energy use. This paper examines how households adapt to chronic grid failure in high-electrification, high-dependence contexts, using Lebanon as a case study. Drawing on original survey data from 1,000 households, we analyze both supply-side coping mechanisms such as diesel generators and solar photovoltaic (PV)-battery systems, and demand-side adaptations, including load shifting and demand suppression. The results reveal a landscape of household responses, where socioeconomic status plays a central role in determining access to backup solutions and the extent of met demand. While diesel generators remain widespread, a transition toward PV-battery systems is observed, especially among financially capable households. However, decentralized self-generation is associated with inefficiencies, including substantial levels of curtailed solar generation. On the demand side, households exhibit reductions in electricity use, leading to distinct consumption profiles depending on the type of backup system employed. These findings highlight the importance of distinguishing between met and unmet demand when assessing energy needs under unreliable supply. The paper contributes to the literature by providing a quantitative characterization of the interaction between self-generation and demand adaptation in a supply-constrained high-electrification context. It also offers empirical demand profiles that incorporate suppressed consumption, addressing a key gap in electricity system planning. From a policy perspective, the results underscore the need to account for unmet demand, address inequities in access to coping technologies, and reduce inefficiencies in decentralized systems.
\end{abstract}




\maketitle

\section{Introduction}\label{sec: intro}

Despite a 92\% global electrification rate in 2025, more than 1.18 billion people still experience energy poverty, lacking access to a sufficient and reliable electricity supply \cite{sdg7_2025, min2024lost}. Failing electric utilities are a growing source of global financial and infrastructural strain, often reflected in frequent and prolonged power outages, high operating costs, and an inability to meet rising demand \cite{WorldBank2024_utilities}. The World Bank estimates that around 40\% of utilities are not financially sustainable. Increasingly, however, this challenge is concentrated in countries where electrification is nearly universal, yet electricity systems remain unable to deliver reliable service.


Countries with historically low electrification rates, such as Ethiopia, Tanzania, Bangladesh, and Nepal, still face unreliability issues, despite important efforts to increase access \cite{berha2026unreliable}. They are characterized by low per capita electricity consumption and reliance on alternative fuels (Table \ref{tab:access_demand}). In contrast, a distinct but less studied group of countries characterized by near-universal electrification but chronic supply shortages faces a different problem. In these contexts, including Lebanon, Ukraine, Iraq, and South Africa,  electricity is deeply embedded in household and economic activities, and shortages occur despite high levels of demand and dependence on electricity \cite{abi2018energy, IEA2024_UkraineEnergySecurity, tahir2025diesel}. These contexts can be described as high-electrification, high-dependence systems under grid failure, where the challenge is not access expansion but the management of persistent supply shortages.

\begin{table}[]
\begin{tabular}{lcccc}
\hline
Country & \begin{tabular}[c]{@{}c@{}}Access rate in\\ 2000 (\%)\cite{WorldBank2024Access} \end{tabular} & \begin{tabular}[c]{@{}c@{}}Access rate in\\ 2020 (\%)\cite{WorldBank2024Access} \end{tabular} & \begin{tabular}[c]{@{}c@{}}Demand in 2020 \\ (MWh/capita)\cite{WorldBank2024Consumptiom} \end{tabular} & \begin{tabular}[c]{@{}c@{}}Electricity  share in energy \\ mix in 2020 (\%)\cite{IEA_WorldEnergyBalances}\end{tabular} \\ \hline
Ethiopia & 12.7 & 51.1 & 0.093 & 2.62 \\
Tanzania & 8.7 & 39.9 & 0.112 & 2.63 \\
Nepal & 29.9 & 94 & 0.257 & 4.24 \\
Bangladesh & 32 & 96.2 & 0.51 & 21.06 \\ \hline
Ukraine & 99.1 & 100 & 2.92 & 20.4 \\
South Africa & 72.4 & 90 & 3.48 & 28.23 \\
Lebanon & 99.3 & 100 & 2.87 & 31.19 \\
Iraq & 96.8 & 100 & 1.18 & 16.4 \\ \hline
\textbf{Global} & \textbf{78.2} & \textbf{91.6} & \textbf{3.26} & \textbf{25.95} \\ \hline 
\end{tabular}
\caption{Electricity access and demand data across different countries}
\label{tab:access_demand}
\end{table}

Households respond to prolonged and often unplanned outages through a range of coping mechanisms, broadly categorized into supply-side and demand-side strategies. While supply-side responses rely on alternative electricity generation, demand-side responses reshape consumption patterns, either by substituting electricity with other energy sources or by suppressing demand altogether \cite{hashemi2022would}. Historically, diesel generators have been the common solution for household electricity generation under grid failure \cite{lawrie2025friend}. Other technologies, such as solar photovoltaic (PV) systems were historically costly, and therefore reserved to industries that could afford the economies of scale. As the price of these systems decreased, households increasingly adopted them as a cheaper alternative to diesel generators \cite{IEA2025, zebra2021review}. In contexts where energy consumption is not as reliant on electricity, or when the economics of backup generators are prohibitively unfavorable, households may choose to energize otherwise electrified activities through alternative fuels \cite{wernecke2024fuel}. The choice of coping mechanism ultimately depends on the consumption habits of the household, and the availability of financing options \cite{dagher2023extreme}.

Understanding these behaviors directly feeds into policy design for the rehabilitation of the electricity sector in these contexts while informing a sustainable energy transition. Studying them gives decision makers insight into the assets available at the consumer level, and the overall shape of household load profiles under unreliable supply. This is particularly relevant as disruptions to electricity supply impose welfare losses on households, with greater impacts expected in countries where reliance on electricity is higher \cite{aweke2022valuing}. Previous literature has predominantly focused on countries at the edge of electrification, with limited quantitative evidence on household behavior in already electrified systems experiencing persistent supply shortages. As a result, the interaction between self-generation and demand adaptation and its implications for system planning remains underexplored particularly in high-dependence contexts.

Lebanon is one such country where both electricity access and demand per capita are high, despite the chronic outages (Table \ref{tab:access_demand}). Taking Lebanon as an example, this paper focuses on the coping mechanisms of countries with a deteriorating grid, where households have high demand. Specifically, we ask: \textit{How do households in high electricity dependence contexts adapt to persistent supply shortages? How do household supply-side coping strategies interact with their demand patterns? What factors drive the household choices, and what are their implications for electricity system design and energy policy?}

In the aftermath of the civil war and persistent geopolitical instability, the Lebanese national electric utility, \textit{Electricit\'e du Liban} (EDL), has consistently failed to meet consumer demand. In response, it rationed supply through scheduled outages, cutting electricity across regions for up to 8 hours per day \cite{abi2018energy}. Following the 2019 economic crisis, fuel import constraints, combined with EDL’s heavy reliance on imported oil (accounting for 95\% of generation), reduced grid availability to as little as two hours per day \cite{olleik2026planning, wehbe2020optimization}. High operating costs, electricity subsidies, low collection rates, and widespread non-payment of bills further accelerated the utility’s financial and operational collapse \cite{ahmad2021distributed}.

Lebanese households and businesses have therefore turned extensively to self-generation. To compensate for grid interruptions, neighborhood diesel generators have long operated as informal microgrids, where private operators generate and distribute electricity under local monopoly conditions \cite{fheili2024off, al2021local, zwickl2025market}. In response to rising costs, the Ministry of Energy and Water has introduced monthly caps on generator tariffs based on a uniform per-kWh pricing scheme, though enforcement remains uneven \cite{ahmad2022dysfunctional}. These diesel-based systems are also concentrated in dense urban areas, raising significant environmental and public health concerns \cite{ahmad2021distributed}. Lebanon thus represents a critical case of a high-electrification system operating under near-continuous grid failure.

More recently, households have increasingly adopted rooftop solar PV-battery systems as a coping strategy against both unreliable grid supply and the high cost of diesel generation \cite{fheili2024off, dbouk2025microgrids}. Prior to the crisis, the central bank supported such investments through subsidized zero-interest loans, but these programs were discontinued following the financial collapse \cite{helou2025informal}. While the declining cost of solar technologies has enabled wider adoption, access remains uneven and largely limited to households with sufficient financial resources and physical space \cite{fheili2024off, dagher2023extreme}. At the policy level, a renewable energy law has been enacted that formally enables feed-in from distributed generation \cite{LCEC_2023}; however, its implementation remains pending. In the absence of grid integration mechanisms, decentralized systems may introduce inefficiencies, including significant levels of unused (curtailed) generation capacity. Overall, reliance on backup generation, whether diesel or solar, has substantially increased household energy expenditures \cite{dagher2023extreme}.
 
Two main streams of studies emerge from the Lebanese context: (1) techno-economic work for the improvement of the sector through centralized \cite{salameh2020economic} and decentralized setups \cite{olleik2026planning, chedid2020optimal}, overlooking the current household-level implications of coping mechanisms and (2) behavioral research relying primarily on limited qualitative evidence: exploratory fieldwork on the experience and behaviors of 10 Lebanese households prior to the 2019 crisis briefly notes the existence of some load shifting behaviors as a result of interruptions on the main grid \cite{abi2018energy}. One of the rare quantitative surveys interviews over 900 households to report on the high incidence of energy poverty in Lebanese households \cite{dagher2023extreme}. 

To answer the research questions, building on the Lebanese case, this study relies on primary data collected through a survey across 1000 Lebanese households, covering some household background, and their consumption and self-generation habits.
This paper contributes to the ongoing discussion about coping mechanisms under grid unreliability, addressing the existing gaps in the literature, in the following aspects:
\begin{itemize}
    \item It introduces a novel dataset of household demand and self-supply in a high-electrification, supply-constrained context.
    \item It provides a quantitative characterization of the joint supply–demand adaptation of households in a high-electrification, supply-constrained context.
    \item It quantifies household demand profiles under unreliable supply, explicitly accounting for both met and unmet demand.
    \item It derives policy-relevant insights for electricity system planning, highlighting inefficiencies in decentralized coping mechanisms and equity implications across socioeconomic groups.
\end{itemize}

The rest of this paper is organized as follows: Section \ref{sec:lit} reviews the available literature on coping mechanisms under grid unreliability. Section \ref{sec:methods} describes the sampling, data collection, and the analytical framework used. Section \ref{sec:results} presents the results of the survey, covering the different coping mechanisms with respect to the relevant factors. Finally, Section \ref{sec:policy} concludes and provides policy insights on the implications of the main findings.

\section{Background} \label{sec:lit}
As the reliability of electric grids deteriorates, consumers have to adapt to prolonged outages. The literature on coping mechanisms discusses the supply and demand-side behaviors adopted by households, often looking at them in isolation from one another, or from other relevant factors. Moreover, the policy implications in these studies often concern countries with low electrification rates, and more rarely those with higher demands.

The adoption of backup generators is a widespread household response in countries under chronic shortages. Diesel generators are a common alternative when the main grid is unavailable. They can be either installed at the household level, or in a neighborhood where they operate in an informal microgrid \cite{lawrie2025friend}. Such setups are prevalent in countries such as Iraq, Lebanon, and Nigeria, in which the main grid has failed due to regional conflicts, dysfunctional governments, and corruption \cite{al2021local, obar2022navigating}. In South Africa, the longstanding load-shedding policies, which implement a schedule of planned outages, have forced households to rely on private diesel generators \cite{lawrie2025friend}. Consumers in Ukraine have likewise resorted to these solutions, driven by the war-related damages on the grid's infrastructure \cite{IEA2024_UkraineEnergySecurity}.
 However, in recent years, households have been increasingly relying on residential solar photovoltaic (PV) systems, which are often more economical when compared to diesel generators \cite{zebra2021review}. In sub-Saharan Africa, households with access to the grid but still suffering from prolonged, unplanned outages significantly adopt solar PV systems \cite{borujeni2022solar}. Households in Ukraine, South Africa, and Lebanon face the same situation \cite{lawrie2025friend}. A review of secondary data identifies 127 factors affecting the decision on adoption of household PV systems, confirming the positive relationship between solar electricity uptake and grid disruptions \cite{shakeel2023solar}. In Bangladesh, a survey of 500 households also reports that solar PV owners are more likely to use their systems as a post-disaster coping mechanism \cite{amin2021solar}.
Another type of supply-side adjustment strategy entails fuel-switching, where households rely on alternative energy sources to perform activities that are otherwise electrified. 
A survey of 6000 households in Nepal examines the fuel-switching decision, and determines that the gender and education level of the head of the household, as well as the household size and location affect the choice of the alternative energy source \cite{acharya2021household}. 

Demand adjustments are a common strategy when households are faced with unreliable access. Load-shifting is the practice of rescheduling electricity-intensive activities to times when affordable supply is available. Load-shifting in the literature is often framed and studied as an intentional result of demand-side management programs or policies, such as price increases or planned load-shedding \cite{usman2022systematic, thakur2016demand}. Metered data from Pakistan shows the effect of an uninterrupted power supply (UPS) system on the rebound peak after electricity is restored \cite{kazmi2024quantifying}. While this behavior is also commonly observed as a result of household PV-adoption in countries with reliable grids, this phenomenon is more rarely discussed in the context of grid-constrained economies \cite{hubert2024laundry}.
When energy is too expensive, or the demand flexibility of consumers is not high enough, households have to resort to load reduction strategies, reducing their overall demand for electricity \cite{hashemi2022would}. A household survey from South Africa shows that respondents tend to engage in electricity-saving activities under load-shedding policies \cite{wiese2024impact}. This reduction can be attributed to fuel-switching, or ceasing certain activities altogether. In these cases, consumption for heating and temperature control is typically the first to be affected \cite{wiese2024impact, ngoma2018households}. A qualitative household survey from France, Spain, Austria, and North Macedonia shows that these mechanisms, particularly around temperature control, are also observed in high-income countries \cite{stojilovska2021out}.

Despite recognizing both types of coping mechanisms, the literature has rarely considered how the supply recovery choice and demand are related. In reality, the listed coping mechanisms are simultaneous strategies adopted by households when faced with chronic shortages and grid unavailability, and therefore affect one another. However, most studies only account for one of these adaptations at a time \cite{carley2022behavioral}. A few surveys have taken a more comprehensive approach to understanding these responses. Fieldwork in Tanzania reports on the different resilience mechanisms adopted by households, such as the use of backup generators, fuel switching, and load-shifting \cite{eledi2023toward}.  A household survey in Zambia finds that over 50\% of respondents used alternative sources for food preparation and preservation, and space lighting. It also reports on the use of backup diesel generators, load-shifting, and load reduction \cite{ngoma2018households}. 
Yet, despite considering multiple coping strategies at once, the available literature lacks a holistic view of the interaction between them. Moreover, these studies primarily concern countries with historically low electrification rates and low electricity demand. In countries with higher reliance on electricity, more insights on the aggregate implications of coping mechanisms at the household level are needed.  


\section{Methodology and data} \label{sec:methods}
\begin{figure}
    \centering
    \includegraphics[width=.8\linewidth]{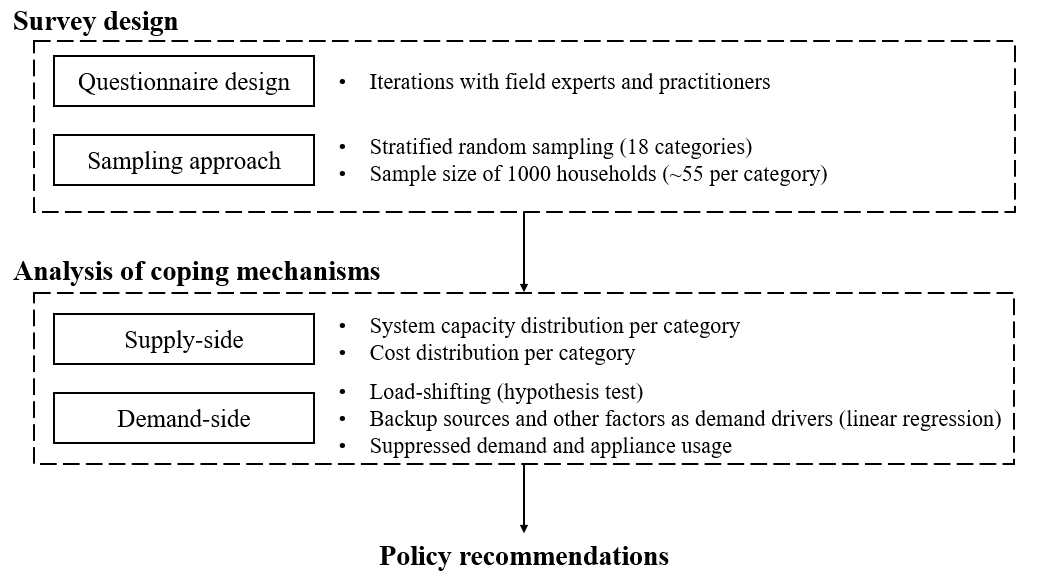}
    \caption{Overview of the research methodology}
    \label{fig:methods_diagram}
\end{figure}

The survey is designed to examine the coping mechanisms of households with high electricity demand under grid unreliability. To do so, the questionnaire is first developed to capture household data on consumption and self-supply habits. The sampling approach captures the different backgrounds of Lebanese households, by randomly selecting participants of pre-determined categories. The collected data is then analyzed to understand the supply-side and demand-side coping mechanisms of households, their factors and magnitudes, eventually leading to policy implications. The methods of the survey design and analysis are summarized in Figure \ref{fig:methods_diagram}, and detailed throughout this section. 

\subsection{Questionnaire}
The survey includes four types of questions, pertaining to (1) general household information, (2) backup electricity sources, (3) typical demand, and (4) unmet demand. 
The general household information collects data on factors known to correlate with the electricity demand \cite{bohlmann2021examining, tokam2025identification}. The geographic division is used to track the coverage of the Lebanese territory.
The majority of Lebanese households use backup electricity sources to mitigate the unreliability of EDL. Households can then be categorized with respect to their backup generation solution: (i) DG-only, relying exclusively on a diesel generator, (ii) PV-only, having installed a PV system, but not connected to a diesel generator, or (iii) hybrid, having both access to a solar system and a diesel generator. 
To identify the typical demand and unmet demand, data on appliance ownership is first collected, and the hourly use of these appliances throughout a representative day is tracked. 
The appliances are categorized into yearly, \textit{i.e.} those for which the usage is constant throughout the year, and seasonal. The survey structure allows for the addition of any major appliances, yearly or seasonal, that are not listed. Seasonal appliances are primarily for temperature control, \textit{i.e.:} cooling and heating. Therefore, four representative days are collected for each household: one for each season of the year, where the demand or unmet demand of each day adds the seasonal difference to the yearly base consumption.
The survey questions were identified through discussions with experts and practitioners, and are summarized in Table \ref{tab: s quest}. The full list of questions is shown in Appendix \ref{app:questions}, and the appliances of interest are listed in Table \ref{tab:apps}.

\begin{table}[]
\begin{tabular}{lll}
\hline
Category & Question & Expected answer type\\ \hline
General information & Socioeconomic status (SES) & Low/ Medium/ High \\
& Number of members in household & Numerical \\
 & Household area (m$^2$) & Numerical \\
 & Geographic division & Categorical\\
 & Number of rooms in household & Numerical \\
 & Solar water heater available & Yes or no \\
 & LED lighting percentage & Numerical (0-100\%) \\ \hline
Electricity sources & Source, consumption and spending & Categorical and numerical\\
 & Barriers to PV installation (if any) & Categorical \\
 & Barriers to solar water heater installation (if any) & Categorical \\
 & PV-battery system size and characteristics (if any) & Categorical \\ \hline
Typical load & Appliances owned & Numerical\\
 & Typical daily and seasonal appliance use & Numerical\\ \hline
Unmet demand & \begin{tabular}[c]{@{}l@{}}Additional daily and seasonal appliance use under 24/7 \\national  grid availability\end{tabular} & Numerical\\ \hline
\end{tabular}
\caption{Survey questions}
\label{tab: s quest}
\end{table}

\begin{table}[]
\centering
\begin{tabular}{cc}
\hline
Yearly appliances & Seasonal appliances \\ \hline
Refrigerator & Inverter AC \\
Freezer & Non-inverter AC \\
Clothes washer & Electric heater \\
Clothes dryer & Irrigation pump \\
Dishwasher & Electric water heater \\
Microwave &  \\
Electric oven &  \\
Electric cooktop &  \\
Water pump &  \\
Lighting &  \\ \hline
\end{tabular}
\caption{Household appliances whose utilization patterns are collected}
\label{tab:apps}
\end{table}

\subsection{Sampling approach}
The unit of analysis in this study is a single household in Lebanon. The selection criterion required that the household only be used for dwelling purposes, excluding any professional activities (e.g.: dental clinics, cafes...). The sample size $n$ is determined using Yamane's formula for sample size calculation \cite{ahmed2024choose} (Equation \ref{eq:sample size}):
\begin{equation}\label{eq:sample size}
    n = \frac{N}{1+N(e^2)}
\end{equation}
where $N$ is the total population size (1.37 M households in Lebanon), and $e$ is the margin of error (3\%). Then, the needed sample size is $\approx 850$ households. To account for potential errors and biases, the final sample size $n$ is rounded up to 1000 households.

\begin{figure}
    \centering
    \includegraphics[width=.7\linewidth]{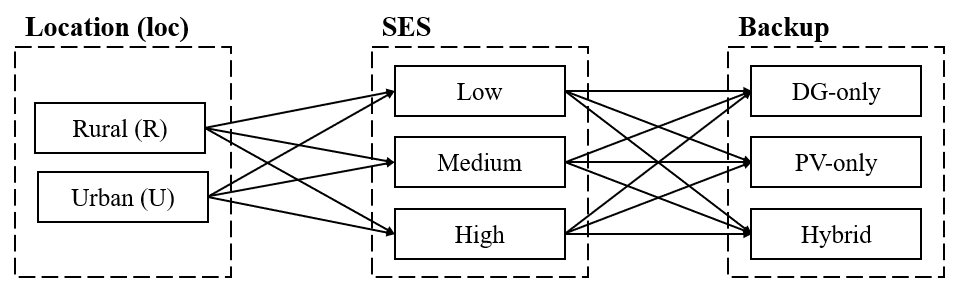}
    \caption{Constituents of the adopted stratification}
    \label{fig:categories}
\end{figure}

The sampling approach relies on stratified random sampling, where each stratum represents a combination of locational, socioeconomic (SES) and backup generation characteristics, listed in Figure \ref{fig:categories}. The sizes of the strata were intended to be equal, to first understand the consumption behavior of each independently, and then compare strata to one another \cite{lohr2021sampling}.
Participant selection for the survey is done via two approaches: random sampling for widely available categories, by asking mayors and municipalities to provide a list of potential participants after getting their consent, and snowballing for categories that are harder to reach, such as residents of remote areas \cite{lohr2021sampling}. Despite its argued potential to yield biased results, snowballing has been a common practice for such hard-to-reach categories \cite{, szymanska2023households}.


\subsection{Analysis framework}
As part of analyzing the collected data, different coping mechanisms are compared along the defined categories. Linear regression is used to understand the correlation between coping mechanisms while controlling for different factors. Different functional forms, built on the same data, are considered, as complementary lenses on the same relationship. This approach is standard in empirical literature, particularly in the case of demand drivers \cite{zarnikau2003functional}.

\paragraph{Lin. Reg. Model} This model corresponds to the linear regression model without any transformations, such that:
\begin{equation}
    y= \beta_0 + \beta_1 x_1 + \beta_2 x_2 + \cdots \beta_k x_k
\end{equation}
where $y$ is the dependent variable, $\beta_0$ is the intercept, and $\beta_1 \cdots \beta_k$ are the coefficients corresponding to the independent variables $x_1 \cdots x_k$. This functional form captures the marginal effects of the independent variables on the dependent one, and is widely used in electricity consumption estimation and forecasting \cite{bianco2009electricity}.

\paragraph{Log Lin. Reg. Model} This model keeps the independent variables intact, but takes the $\log$ of the dependent variable on the left-hand side (LHS), such that:
\begin{equation}
    \log(y)= \beta_0 + \beta_1 x_1 + \beta_2 x_2 + \cdots \beta_k x_k
\end{equation}
In this case, the coefficient of the independent variables are semi-elasticities, explaining the percentage change in $y$ after a unit increase in $x$. 

\paragraph{Log-Log Reg. Model} This model applies a $\log$ transformation on both of the dependent variables, and the continuous independent variables, such that:
\begin{equation}
    \log(y)= \beta_0 + \beta_1 \log(x_1) + \beta_2 \log(x_2) + \cdots + \beta_i \log(x_i) + \beta_j x_j + \cdots+  \beta_k x_k
\end{equation}
where $x_1$...$x_i$ are continuous variables, and $x_j$...$x_k$ are binary ones. In this form, the coefficients of the logged variables can be interpreted as elasticities \cite{bianco2009electricity, filippini2011short}.

In this paper, linear regression is used for inference purposes, as opposed to prediction. The proposed framework focuses on estimating the direction and magnitude of the relationships between electricity demand and its driver. The simplicity of the linear regression interpretation is therefore useful, despite the existence of more advanced tools \cite{yildiz2017review}.

\section{Results and discussion} \label{sec:results}
\begin{figure}
    \centering
    \includegraphics[width=0.5\linewidth]{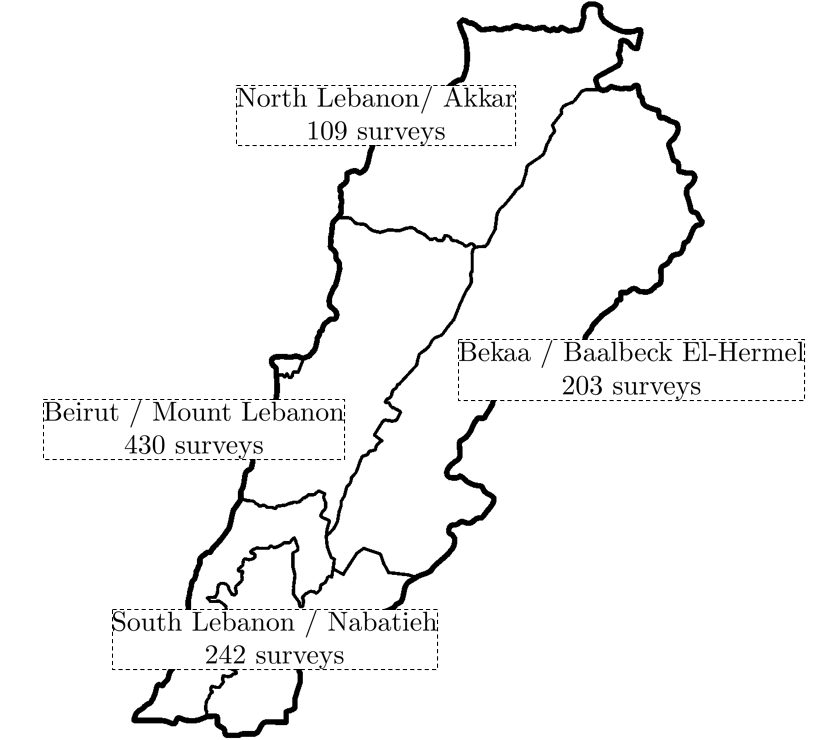}
    \caption{Survey distribution across Lebanon}
    \label{fig:leb_map}
\end{figure}

\begin{figure}
    \centering
    \includegraphics[width=0.75\linewidth]{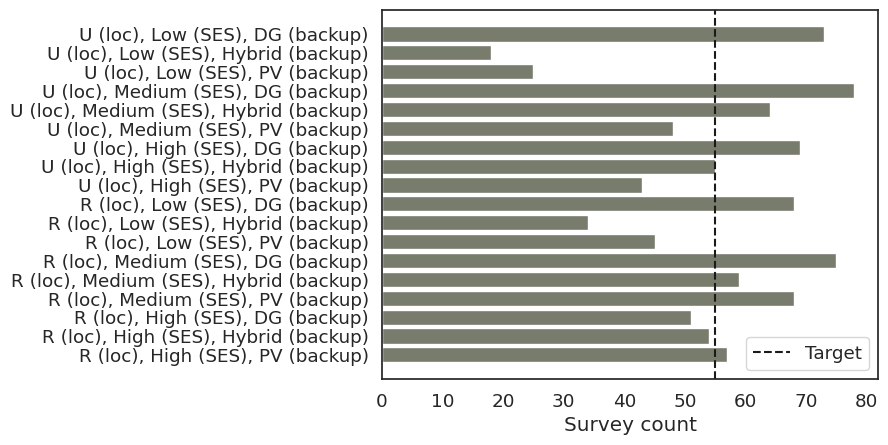}
    \caption{Number of surveys in each category}
    \label{fig:obs_cats}
\end{figure}

Data was collected for 1010 participating households, between the months of March and July of 2025. The final dataset contains 984 surveys, obtained after excluding some surveys that do not fit the selection criteria, or have some reporting irregularities, eliminating less than 3\% of the collected data. The distribution of collected surveys across the Lebanese territory is shown in Figure \ref{fig:leb_map}. Collection efforts aim at ensuring a nationally distributed, stratified, and analytically balanced survey. However, due to the lack of reliable statistics on the household geographic distribution across Lebanon, national representativeness of the survey cannot be claimed. Figure \ref{fig:obs_cats} shows the number of surveys collected for every category. Low-SES households with a hybrid backup setup have noticeably low collection rates. However, this category is relatively uncommon, likely due to the high costs required for hybrid configurations, which tends to be associated with higher-SES households. The number of observations of different questions are summarized in Table \ref{tab: obs quest}. Out of the 984 surveys, 58\% of households have a solar PV-system, 73\% have access to a diesel generator (but only 23\% report on the consumption in kWh of diesel-based generation), and over 96\% maintain a connection to EDL. Finally, 65.8\% of households express some unmet demand. The complete dataset is available online \cite{Dbouk2025}. 

\begin{table}[]
\begin{tabular}{lc}
\hline
Question & Number Observed \\ \hline
Number of members in household & 984 \\
Household area (m$^2$) & 984\\
Geographic division & 984\\
Number of rooms in household & 984\\
Solar water heater available (y/n) & 983 \\
LED percentage (0-100) & 984 \\
PV ownership & 570\\
\begin{tabular}[c]{@{}l@{}}Household DG ownership/with report on spending/ \\ and consumption respectively\end{tabular}  & 73/67/3\\
\begin{tabular}[c]{@{}l@{}}Neighborhood DG subscription/ with report on spending/ \\ and consumption respectively\end{tabular} & 641/635/162 \\
EDL subscription & 948 \\
Typical load & 984 \\ 
Unmet demand & 648\\\hline
\end{tabular}
\caption{Number of observed answers}
\label{tab: obs quest}
\end{table}

\subsection{Generation coping strategies}
Lebanese households have relied on backup generation solutions for decades. They have historically included neighborhood diesel generators, but recently, households have been increasingly adopting rooftop solar PV-battery systems. 

\begin{figure}
    \centering
    \includegraphics[width=0.5\linewidth]{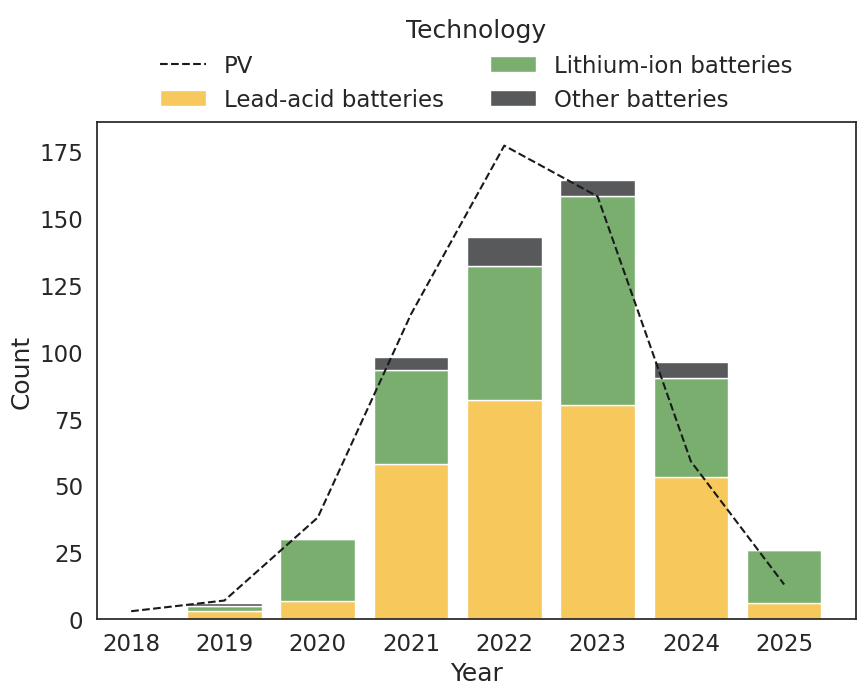}
    \caption{PV installation times and battery installation times and categories}
    \label{fig:bat_dist}
\end{figure}

Prior to the 2020 crisis, households adopting PV-battery system were doing so due to their techno-economic advantages, despite their high investment costs. Therefore, the share of lithium-ion batteries was initially higher (Figure \ref{fig:bat_dist}). However, following the collapse of EDL in 2020, the urgent need for alternative electricity sources pushed households to adopt PV systems rapidly, often favoring the lower upfront cost of lead-acid batteries. In recent years, as the cost of lithium-ion technology has declined, their share has increased again, reflecting their longer lifespans and superior performance. Among the 570 surveyed households owning PV systems, the earliest installations date back to 2018, with a median installation period between 2022 and 2023, coinciding with the peak of the crisis.

\begin{figure}
    \centering
    \includegraphics[width=0.7\linewidth]{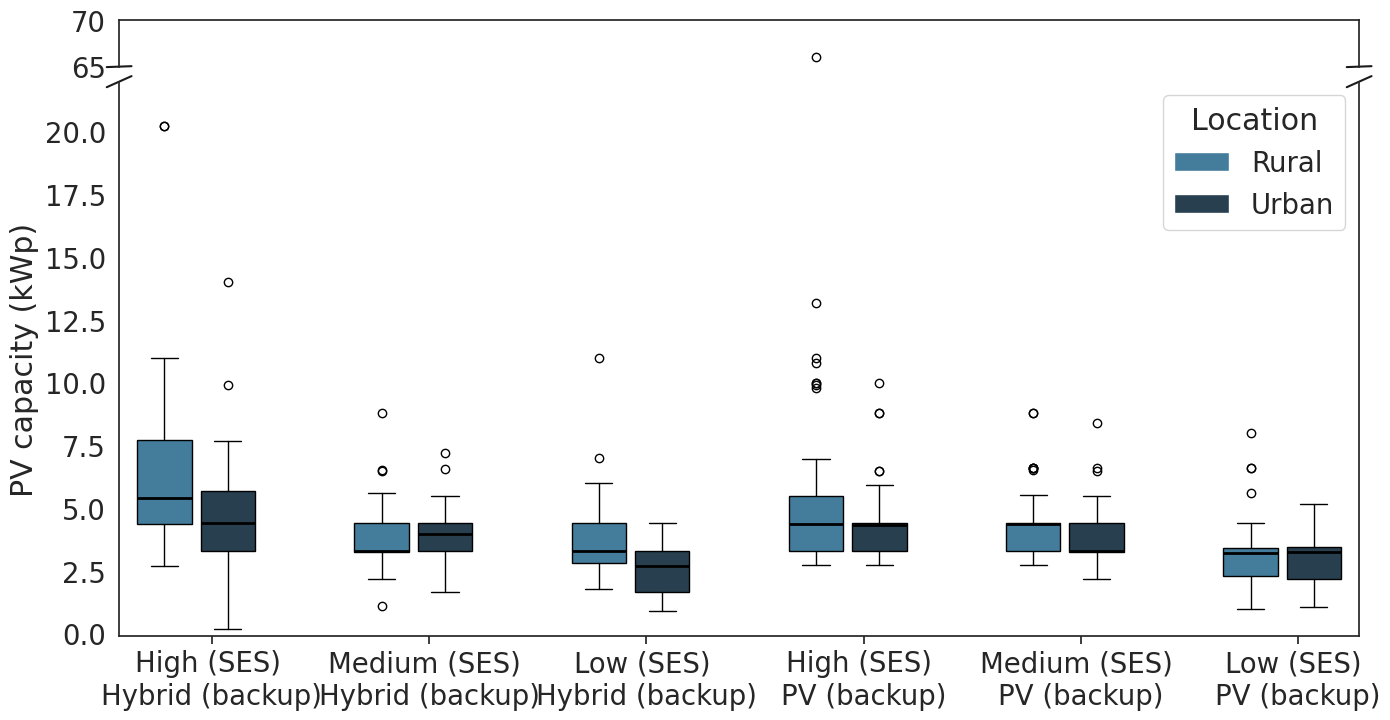}
    \caption{Solar PV capacity distribution}
    \label{fig:pv_dists}
\end{figure}

\begin{figure}
    \centering
    \includegraphics[width=0.5\linewidth]{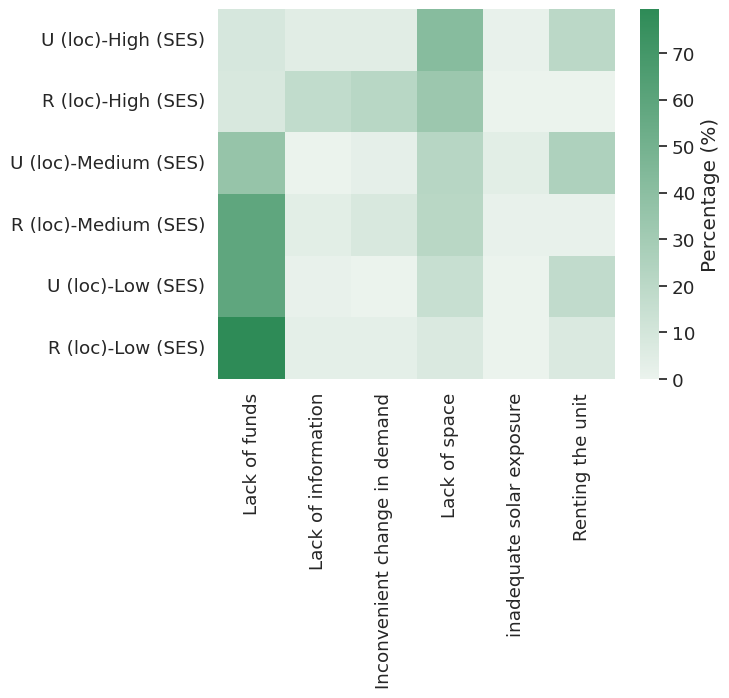}
    \caption{Barriers to household PV adoption}
    \label{fig:no_pv_hm}
\end{figure}

Figure \ref{fig:pv_dists} shows the distribution of these system sizes throughout different categories, grouped by SES and backup generation source. 
Households with higher SES tend to have larger systems, particularly in rural areas, where there is more space for solar PV installation. The median PV capacity is 22\% higher for households with high SES compared to the medium category, and 35\% higher compared to the low category, with the battery capacity following this trend. 
Despite increasing affordability and the potential for improved energy security, several barriers to household PV adoption persist. Among lower-SES households, limited financial resources remain the primary obstacle. In the case of higher-SES households, limited available space is the dominant constraint, particularly in urban areas (Figure \ref{fig:no_pv_hm}).

\begin{figure}
    \centering
    \includegraphics[width=.7\linewidth]{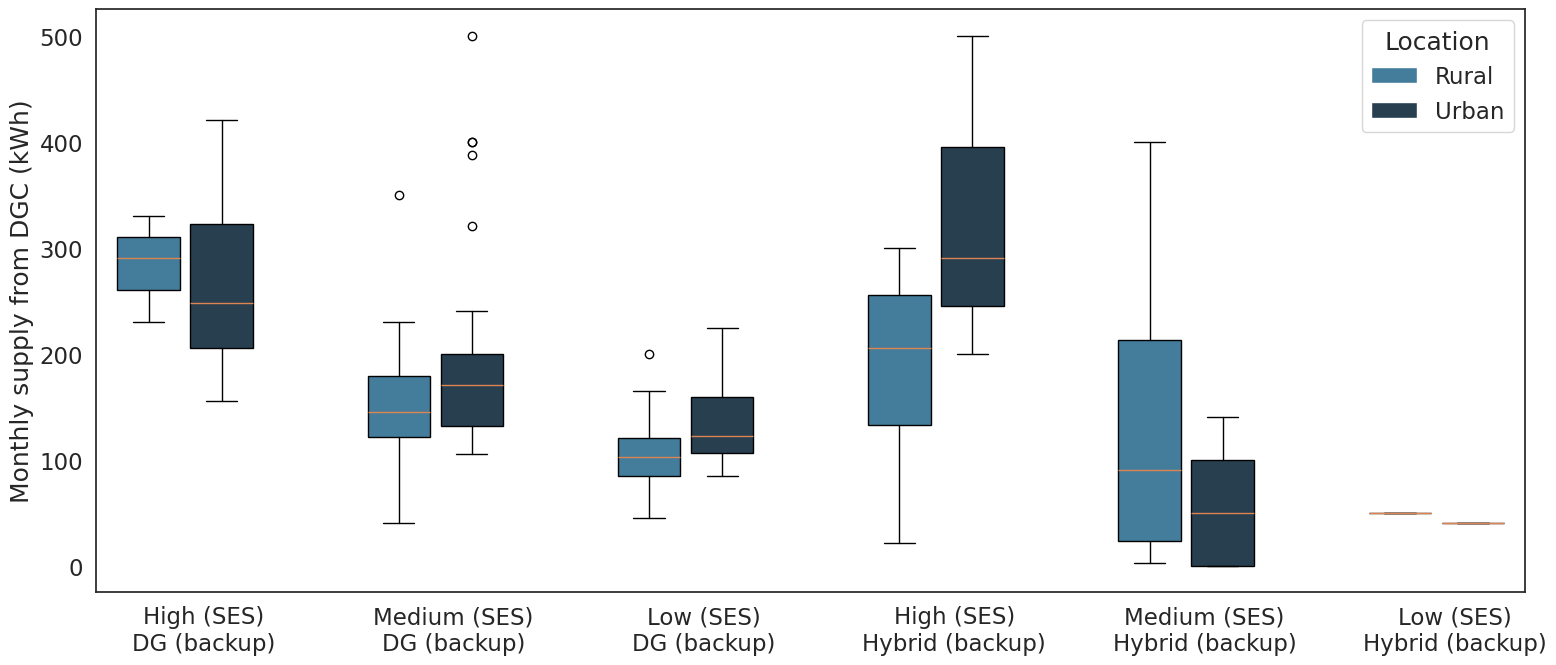}
    \caption{Distribution of monthly supply from diesel generator per household}
    \label{fig:dg dist}
\end{figure}

Despite the widespread adoption of residential PV systems, and partly due to the barriers shown in Figure \ref{fig:no_pv_hm}, neighborhood diesel generators remain a central source of electricity for Lebanese consumers. Figure \ref{fig:dg dist} reports on the monthly supply in kWh billed by the DGCs per household. This consumption highlights the dependency of households on diesel generators to adapt to EDL's outages. Two observations can be confirmed: (1) the consumption from diesel-connected households increases with SES; and (2) hybrid households (who have access to a PV system) are generally less reliant on diesel-based generation. 
The minimal spread of the low (SES) - hybrid (backup) category is due to the limited number of observations in this category, of which only a subset (1 in an urban setting, and 2 in a rural one reported monthly consumption.

\subsection{Demand coping strategies}
As part of their coping behaviors in response to the national grid's failure, Lebanese consumers change their electricity demand, depending on multiple factors. These mechanisms include load shifting, and changes in the volume of demand.

\subsubsection{Load shifting for PV-owners}
\begin{figure}
    \centering
    \includegraphics[width=0.5\linewidth]{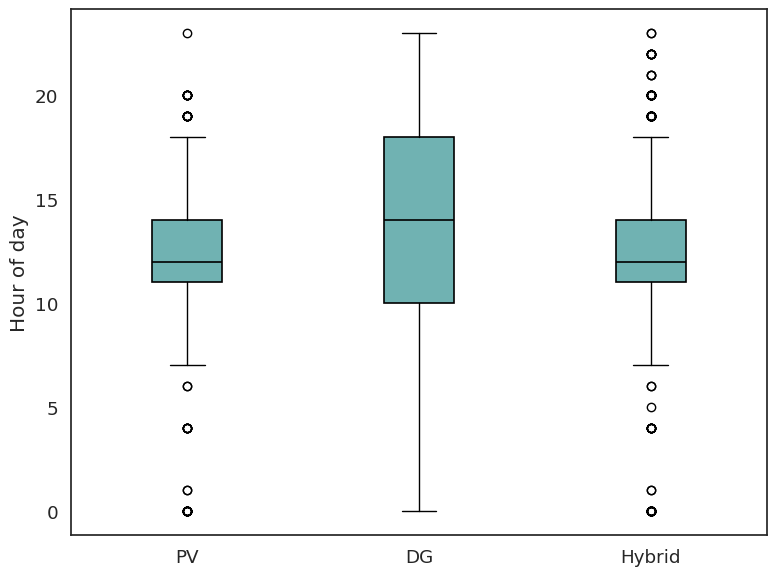}
    \caption{Distribution of hours of peak demand by backup supply source}
    \label{fig:peak_dem}
\end{figure}

A commonly observed behavioral response to residential solar PV adoption, across both developed and developing countries, is the shifting of electricity demand toward earlier hours of the day, aligning with periods of peak solar generation \cite{hubert2024laundry}. As illustrated in Figure \ref{fig:peak_dem}, households equipped with PV-battery systems (whether PV-only or hybrid) exhibit an earlier median hour of peak demand during the day compared to households that rely solely on diesel generators.

To confirm this difference, a hypothesis test compares the average hours of peak demand for PV-owners and non-PV owners. Two one-tailed independent t-tests are considered, one between households from the ``PV''  and ``Diesel generator'' backup categories, and another one comparing the ``Hybrid'' and ``Diesel generator'' backup categories. In each case, $\mu_{PV}$ and $\mu_{noPV}$ are the mean peak demand hour during a day of PV owners and non-PV owners respectively:
\begin{align}
    &H_0\text{ (null)}: \mu_{PV} \geq \mu_{noPV}\\
    &H_1: \mu_{PV} < \mu_{noPV}
\end{align}

The one-tailed independent t-test (results in Table \ref{tab:peak T}) confirms that PV owners exhibit a statistically significant shift in peak demand. This result includes hybrid households who, despite the flexibility of diesel generators, still choose to shift their peak consumption to earlier hours, matching peak sunlight at around 1 PM.

\begin{table}[]
    \centering
    \begin{tabular}{cccc} \hline
      Test   & T-statistic & P-value & Recommendation\\ \hline
       PV vs DG  & -4.034 & 2.818 $\times 10^{-5}$ & \begin{tabular}[c]{@{}l@{}} Reject $H_0$, $\mu_{PV}$ is significantly \\  smaller than $\mu_{noPV}$\end{tabular}\\
       Hybrid vs DG & -4.040 & 2.742 $\times 10^{-5}$ &\begin{tabular}[c]{@{}l@{}} Reject $H_0$, $\mu_{PV}$ is significantly \\ smaller than $\mu_{noPV}$\end{tabular}\\ \hline
    \end{tabular}
    \caption{Student T-test result for mean time of peak demand comparison}
    \label{tab:peak T}
\end{table}

\subsubsection{Met demand} \label{tot_dem}
To understand the change in overall met demand, while accounting for the different confounding variables beyond the backup generators, different linear regression models are studied. In each model, the LHS variable is the daily demand in kWh (or, where it applies, the $\log$ of the daily demand). The independent variables are the different factors suspected to affect demand, inspired by the literature \cite{tokam2025identification}. Two sets of variables are considered in each formulation:
\begin{itemize}
    \item Set of ``base'' variables: these are the variables used for stratification (location, SES, and type of backup generation), as well as the season used in the demand estimation.
    \item Set of extended variables: these are the base variables, in addition to other factors known to affect the demand \cite{tokam2025identification}. These factors include the size of the household (number of rooms, area, and number of members), as well as the existence of some widespread electricity-saving appliances (solar water heater and LED lighting). 
\end{itemize}
Given that all base variables are binary, the \textit{Log-Log} formulation for the base variables becomes the same as the corresponding \textit{Log-Lin} one. Therefore only the full set of variables is considered for this model.
The regression results of each model are summarized in Table \ref{tab:reg_res}.

\begin{table}[]
\centering
\footnotesize
\begin{tabular}{lccccc}
\hline
 & \multicolumn{5}{c}{Coefficients} \\ \hline
\multirow{2}{*}{Variable} & \multicolumn{2}{c}{Lin. Reg.} & \multicolumn{2}{c}{Log Lin. Reg.} & Log-Log Lin. Reg. \\ \cline{2-6} 
 & Base & Extended & Base & Extended & Extended \\ \hline
Intercept & 13.9860*** & 14.4847*** & 2.3834*** & 2.3929*** & 2.714*** \\
SES (Low) & -8.9119*** & -8.0932*** & -0.6446*** & -0.5973*** & -0.5940*** \\
SES (Medium) & -4.1718*** & -3.6098*** & -0.2336*** & -0.2097*** & -0.2111*** \\
Source (PV) & -2.7287*** & -2.7026*** & -0.1747*** & -0.2045*** & -0.2068*** \\
Source (Hybrid) & 0.3985 & -0.0374 & -0.0115 & -0.0533* & -0.051* \\
Location (Urban) & 0.8084* & 0.7153* & 0.0222 & 0.0292 & 0.0233 \\
Season (Spring) & 0.5098 & 0.5098 & -0.0018 & -0.0018 & -0.0018 \\
Season (Summer) & 11.4563*** & 11.4563*** & 0.6849*** & 0.6849*** & 0.6849*** \\
Season (Winter) & 6.3379*** & 6.3379*** & 0.4109*** & 0.4109*** & 0.4109*** \\
Number of rooms &  & 1.4031*** &  & 0.1068*** & 0.4035*** \\
Solar water heater (Yes) &  & -1.535*** &  & -0.0281 & 0.0339 \\
Number of members &  & 0.3798* &  & 0.0623*** & 0.146*** \\
Area &  & -0.3681 &  & -0.0688*** & -0.2089*** \\
LED percentage &  & -0.2168 &  & 0.0047 & -0.0142* \\ \hline
$R^2$ & 0.228 & 0.239 & 0.248 & 0.264 & 0.268 \\
Adjusted $R^2$ & 0.226 & 0.237 & 0.246 & 0.261 & 0.266 \\
F-statistic & 144.7 & 94.9 & 161.8 & 108.1 & 110.7 \\ \hline
\end{tabular}
\caption{Regression results for the sets of base and extended variables, for different models. Significance levels: * $p<0.10$, ** $p<0.01$, *** $p<0.001$.}
\label{tab:reg_res}
\end{table}

The sign consistency of the statistically significant coefficients between different models confirms that: (1) SES is positively correlated to the overall demand, 
(2) households have a higher demand for electricity in the summer and winter, due to the use of appliances for thermal comfort (air conditioning, heaters...), and (3) PV-ownership reduces electric consumption, albeit more moderately when the household also has access to a diesel generator. This last result contradicts the solar rebound effects observed in developed countries \cite{nguyen2024solar}, highlighting the scarcity conditions pushing towards residential PV adoption, as opposed to cost efficiency and environmental concern. Overall, the consistency in $R^2$ shows that the base variables, \textit{i.e.:} SES, source, location, and season are as good at capturing the demand patterns of households as the extended set.

\subsubsection{Unmet demand}
\begin{figure}
    \centering
    \begin{subfigure}[b]{0.49\textwidth}
        \centering
        \includegraphics[width=\textwidth]{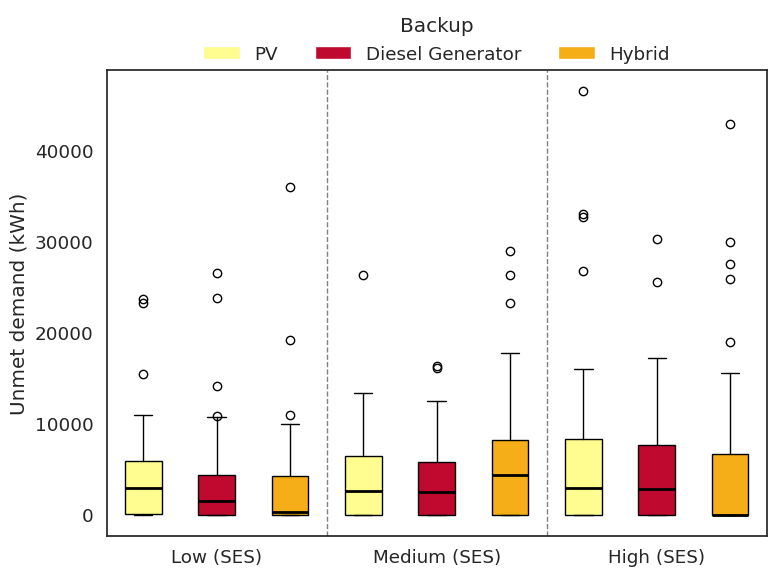}
        \caption{Unmet demand}
        \label{fig:u_demand_kwh}
    \end{subfigure}
    \begin{subfigure}[b]{0.49\textwidth}
        \centering
        \includegraphics[width=\textwidth]{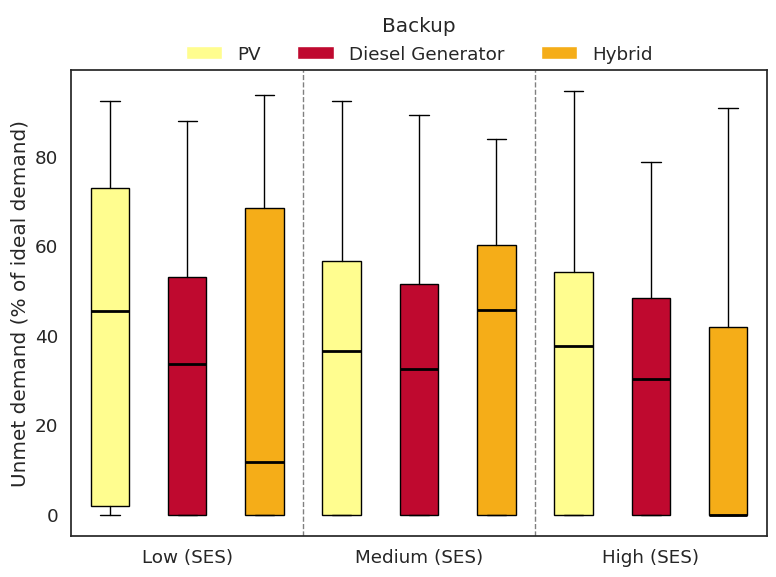}
        \caption{Unmet demand as a percentage of ideal demand}
        \label{fig:u_demand_perc}
    \end{subfigure}
    \caption{Box plots of yearly unmet demand distributions}
    \label{fig:u_demand}
\end{figure}

As part of adapting to the chronic shortages through reshaping their demand, households often end up suppressing part of it. As reported in Table \ref{tab: obs quest}, over 65\% of respondents expressed some level of unmet demand that would be resolved by the availability of the national grid at the current prices (as of 2025). The level of unmet demand varies by category, with its distribution shown in Figure \ref{fig:u_demand}. Except for the medium-SES hybrid results, \ref{fig:u_demand_perc} shows that the unmet demand is highest for PV-only households, followed by diesel-only, while hybrid households exhibit the lowest percentage of unmet demand. Moreover, the higher the SES, the lower the unmet demand percentage across all backup options. 
The unmet demand is affected simultaneously by different factors, including the physical constraints on the feasibility of back-up generation expansion (space limitations for PV installation, and capacity limit from DGCs), and the availability of funds to do so. The effect of these factors on households are further detailed in Section \ref{sec:rep_prof}.

\subsubsection{Appliance ownership and usage}
\begin{figure}
    \centering
        \begin{subfigure}[b]{0.75\textwidth}
    \centering
        \includegraphics[width=\linewidth]{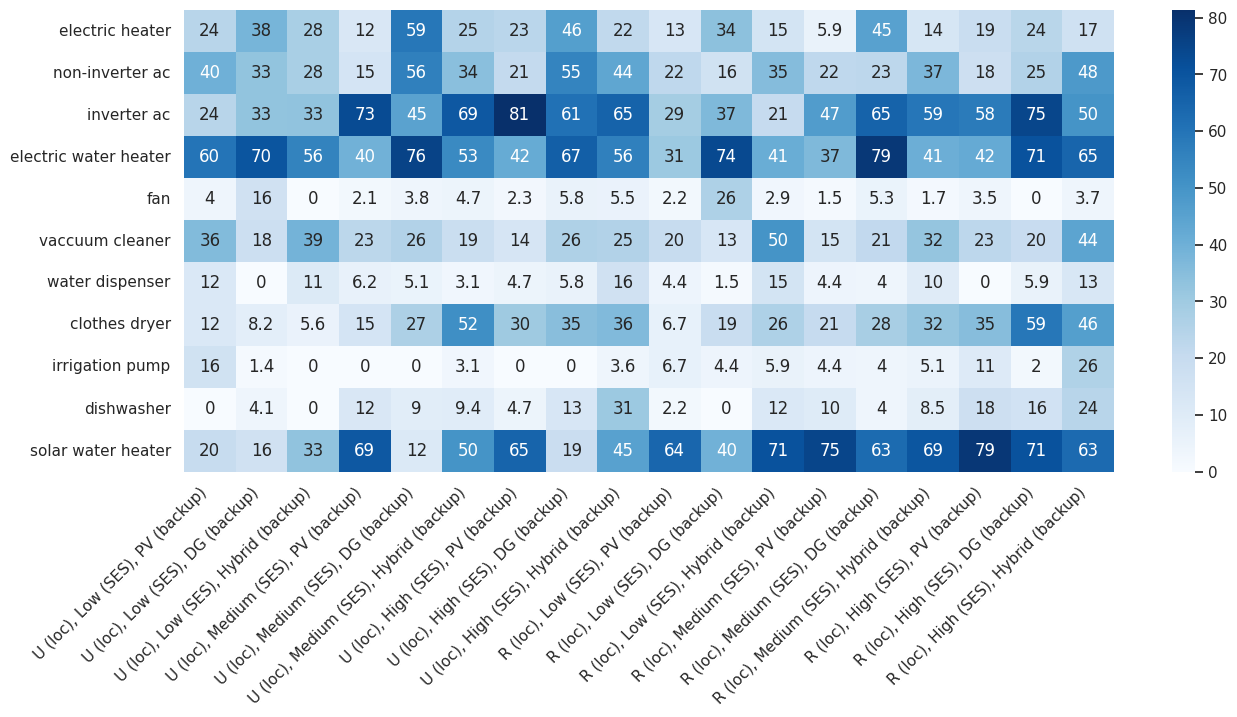}
        \caption{Share of households owning the appliances}
        \label{fig:app_own}
    \end{subfigure}
    \begin{subfigure}[b]{0.75\textwidth}
    \centering
        \includegraphics[width=\linewidth]{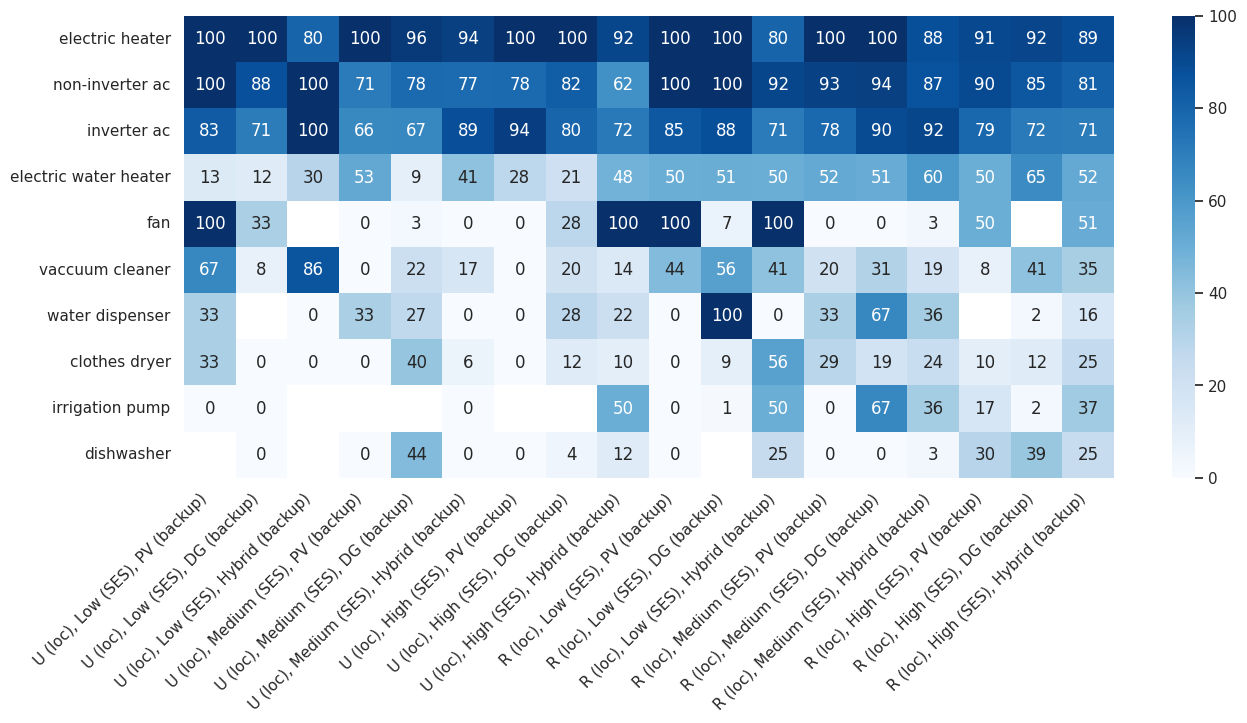}
        \caption{Least used appliances}
        \label{fig:app_diff}
    \end{subfigure}
    \caption{Heatmaps of appliance ownership and use}
    \label{fig:app}
\end{figure}

In addition to behavioral adaptations such as load shifting and reduction, households also adjust their electricity consumption through the acquisition and use of appliances. Figure \ref{fig:app} illustrates how ownership patterns change for the different considered categories, showing the percentage of households owning an appliance (Figure \ref{fig:app_own}), and the percentage of those households leaving it unused \ref{fig:app_diff}. For easier comparison, both heatmaps are sorted by the appliances that are used the least.

The obtained results concur with the literature on demand reduction on temperature control. The appliances that are owned but used the least are the heating and cooling appliances, as seen in the top rows of Figure \ref{fig:app_diff}. This flexibility in thermal comfort is possible due to the moderate weather of Lebanon, where the number of heating degree days and cooling demand are 135 and 130 specifically, in Beirut throughout 2024 \cite{pfenninger2016long, staffell2016using}. 

\subsection{Macro-level insights} \label{sec:policy}
Looking forward, planning and policy-making should consider the entrenched household adaptations emerging from decades of chronic outages. This includes understanding the new shape of the demand of households, evaluating the extent of excess PV generation potential that could be used, and assessing how these adaptations affect energy poverty.

\subsubsection{Representative profiles} \label{sec:rep_prof}
\begin{figure}
    \centering
    \includegraphics[width=0.5\linewidth]{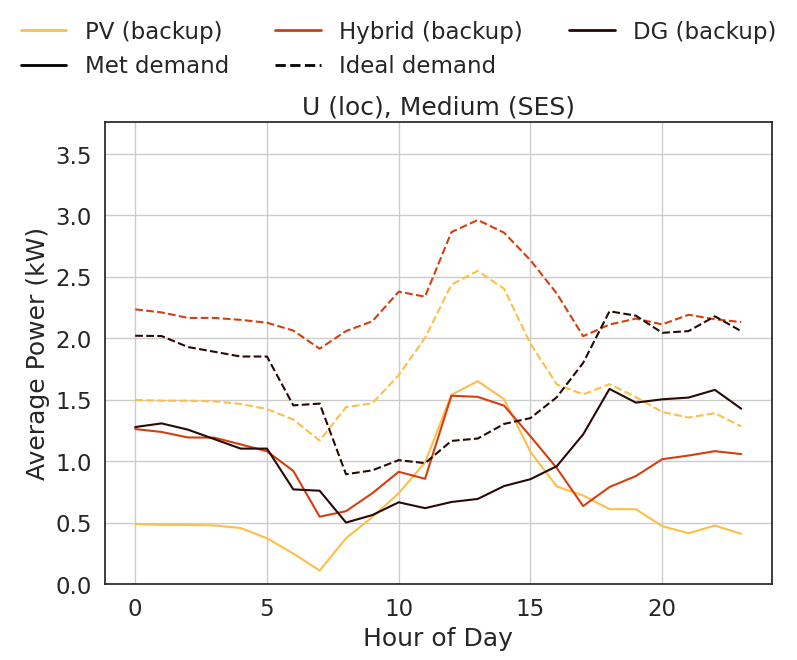}
    \caption{Representative load profiles for urban, medium SES households in summer}
    \label{fig:rep_prof1}
\end{figure}

The representative profiles of every category can be derived, accounting for the current demand and additional unmet demand. Understanding them provides essential insights for policy design around pricing, or the rehabilitation of the electricity grid.
After confirming the significance of the chosen strata criteria through the regressions (Section \ref{tot_dem}), the representative load profiles and ideal profiles (including unmet demand) of each category are obtained. Figure \ref{fig:rep_prof1} shows the profiles of urban, medium SES households in summer for different backup sources. All profiles across seasons and strata are shown in Appendix \ref{app:app_rep_prof}.

 These profiles confirm the previously discussed load-shifting and load-reduction behaviors of PV-owners. They also show that including any additional consumption from EDL to obtain the ideal demand does not change the shape of the profiles. The unmet demand itself can be partly explained through the dynamics in Table \ref{tab:UD}. High SES households might express unmet demand if they only have one backup generation solution. In the case of PV-only households, additional generation from EDL or the DGC could be affordable, but the DGC might not allow their connection to the informal microgrid \cite{zwickl2025market}. Similarly, DG-only households can afford additional generation, but are either limited by their subscription to the DGC, or by the physical constraints on the installation of a PV-battery system. 36\%  of the non-PV owners respondents in medium and high SES households mention at least one of the following barriers to PV adoption: lack of space, inadequate solar exposure, or conflicts with neighborhood agreements. In the case of low SES households, who cannot afford backup generation expansion (irrespective of physical feasibility), unmet demand only arises if the EDL tariffs are lower than the levelized cost of electricity of additional PV capacity, or the per-kWh cost of DGC generation. Otherwise, the household cannot afford any additional load on the national grid if reliability is restored, and therefore has no unmet demand as defined in this paper.

\begin{table}[]
\small
\begin{tabular}{ccl}
\hline
\multicolumn{2}{c}{Backup generation expansion} & \multicolumn{1}{c}{\multirow{2}{*}{Effect}} \\ [0.2cm]
\multicolumn{1}{c}{\begin{tabular}[c]{@{}c@{}}Physical \\ feasibility\end{tabular}} & \multicolumn{1}{c}{\begin{tabular}[c]{@{}c@{}}Availability \\ of funds\end{tabular}} & \multicolumn{1}{c}{} \\ \hline
 \checkmark & \checkmark & \begin{tabular}[c]{@{}l@{}}Households in these situations  exhibit the lowest levels of unmet demand, \\ notably, hybrid households with a   higher SES.\end{tabular} \\ \hline
\XSolidBrush & \checkmark & \begin{tabular}[c]{@{}l@{}}This situation often concerns    higher-SES households with a single back-up \\ generator. PV-only and diesel generator households in these cases can afford\\ additional spending on EDL electricity if available, but cannot increase their \\ backup capacity. They therefore might express unmet demand. \end{tabular} \\\hline
\checkmark or \XSolidBrush & \XSolidBrush & \begin{tabular}[c]{@{}l@{}}This set of constraints mainly affects lower SES-households. In this \\ situation the level of unmet demand is dependent on the unit economics of the \\ available generation technologies. \end{tabular} \\ \hline
\end{tabular}%
\caption{Unmet demand dynamics}
\label{tab:UD}
\end{table}

\subsubsection{Wasted PV excess}
\begin{figure}
    \centering
    \begin{subfigure}[b]{.49\textwidth}
        \includegraphics[width=\linewidth]{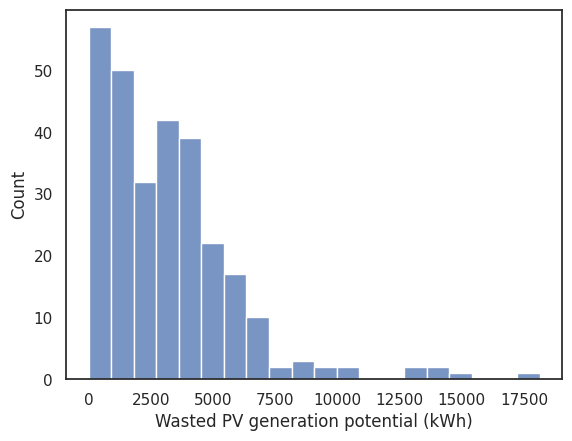}
        \caption{Energy}
        \label{fig:wasted_pv_ene}
    \end{subfigure}
    \begin{subfigure}[b]{.49\textwidth}
        \includegraphics[width=\linewidth]{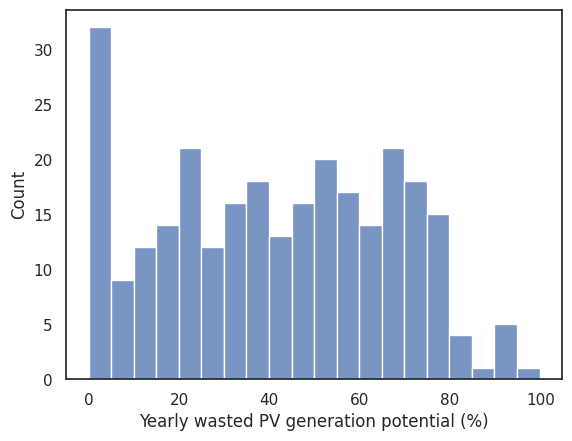}
        \caption{Percentage}
        \label{fig:wasted_pv_perc}
    \end{subfigure}
    \caption{Wasted household PV generation potential}
    \label{fig:wasted_pv}
\end{figure}

The total yearly generation of PV-only households can be found from the collected data on PV system capacity and the average solar capacity factor in Lebanon, estimated at 0.21 \cite{boukather2023re}. Knowing the demand of households, the wasted PV generation potential can be found. Figure \ref{fig:wasted_pv} shows this distribution for PV-only households. While the exact excess depends on the availability of EDL, the net difference between demand and supply from PV suggests that PV-only households waste at least 41\% of their generation potential on average, amounting to over 2.4 MWh per year per household. This excess can be used in the design of a policy promoting PV-owner exchange with the grid.

\subsubsection{Energy poverty}
\begin{figure}
    \centering
    \includegraphics[width=\linewidth]{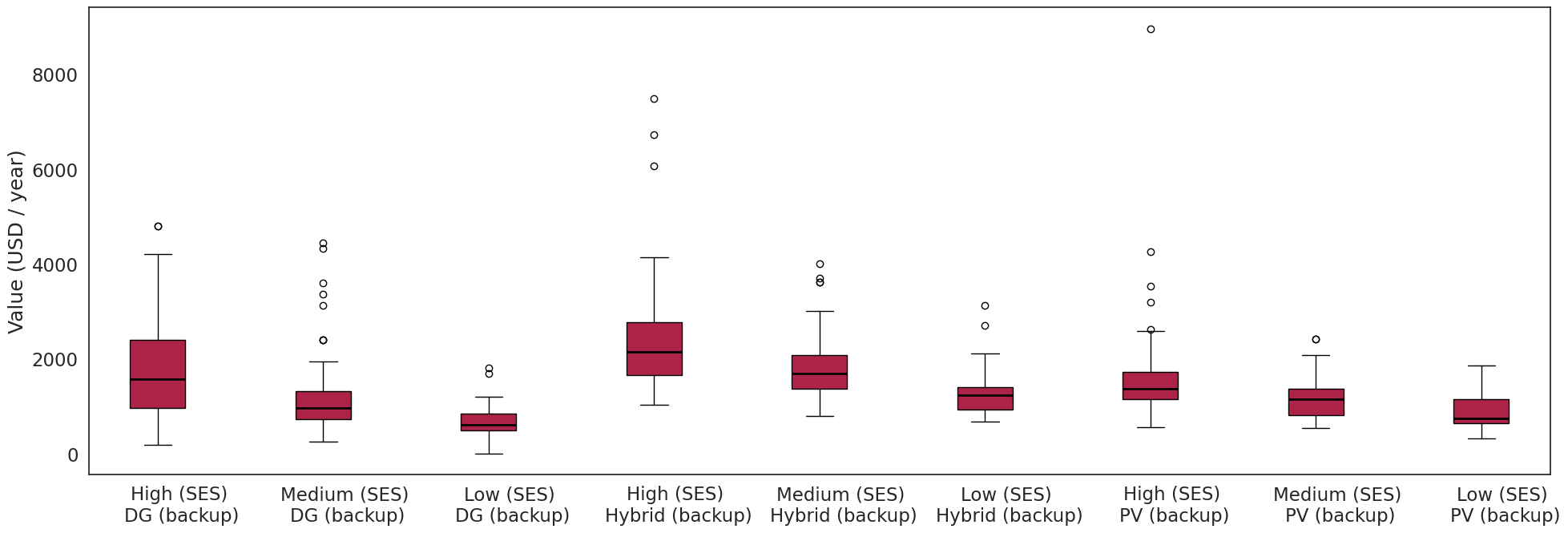}
    \caption{Distribution of annual costs}
    \label{fig:costs_y}
\end{figure}

Backup technologies impose additional costs on households beyond EDL tariffs. These costs are captured through survey data on neighborhood diesel generator bills and PV–battery system sizes, from which annual expenditures are estimated. Figure \ref{fig:costs_y} shows the distribution of these costs across the different categories, with higher SES associated with higher costs.

National statistics on the income of Lebanese households are rare. Informal sources place the average yearly household income between 3,600 USD and 6,100 USD \cite{TheBeiruter_2025, wage.is_2026}. The average yearly expenditure on backup generation alone therefore represents 22\% to 38\% of the average yearly household gross income. 

\section{Conclusions and policy implications} \label{sec:policy}

This paper investigates household coping mechanisms under chronic electricity supply shortages, focusing on Lebanon as a representative case of high-electrification, high-dependence systems with deteriorating grid performance.
Using original survey data from 1,000 households, the study analyzed both supply-side and demand-side adaptations, and examined how these responses interact with socioeconomic characteristics and system constraints.

The findings reveal a diversified landscape of generation coping mechanisms, with households relying on three main configurations: diesel generators, solar PV-battery systems, and hybrid combinations. Households installing PV-battery systems are increasingly shifting towards lithium-ion battery technologies reflecting their improving affordability and performance. A striking result is the extent of curtailed solar generation, which exceeds 41\% of potential production among PV-owning households, highlighting significant inefficiencies in the current decentralized system.

On the demand side, the analysis emphasizes that understanding electricity consumption under unreliable supply requires distinguishing between met and unmet demand. Socioeconomic status plays a central role in shaping the level of met demand, with higher-income households better able to maintain consumption levels through backup solutions. Moreover, demand profiles vary significantly depending on the type of backup system used, particularly in terms of peak load behavior. The study also documents both voluntary and involuntary demand reductions, including the partial or complete suspension of certain electrical uses, which substantially reshape household load profiles. These findings underscore the importance of accounting for suppressed demand in electricity system planning and demand forecasting.

The results carry several important policy implications. First, electricity system planning in contexts of chronic shortages should move beyond observed consumption and explicitly incorporate unmet demand that a reliable national grid should also expect to satisfy. By providing empirical estimates of both met and unmet demand, this study contributes to a more accurate representation of actual energy needs.

Second, the strong link between socioeconomic status and coping capacity raises concerns about equity. Addressing these disparities requires targeted policy interventions, such as reintroducing subsidized or low-interest financing schemes for residential solar PV systems, and regulating the pricing structure of diesel generator services through progressive tariff mechanisms rather than uniform pricing.

Third, the high level of curtailed solar generation points to substantial inefficiencies and missed opportunities for system optimization. In the short term, enabling localized energy exchange, particularly between solar-equipped households and diesel generator-based microgrids, could improve efficiency, reduce costs, and lower emissions. In the longer term, operationalizing the existing renewable energy law through a functional feed-in tariff scheme, in line with existing renewable energy legislation, would allow surplus distributed generation to be integrated into the national grid. Such measures, however, are contingent on parallel investments in grid rehabilitation and regulatory reform.

While the survey covers approximately 1,000 households that are spread across the Lebanese geography, it does not necessarily constitute a nationally representative sample. Future research should aim to validate these findings using potentially larger datasets, as well as higher-resolution consumption data. In addition, longitudinal analyses would help capture the dynamic evolution of coping strategies as economic conditions, technology costs, and policy frameworks change over time. Despite these limitations, the findings provide robust insights into household behavior under extreme supply constraints and offer a valuable empirical basis for designing more resilient and equitable electricity systems. These insights are directly relevant to a growing number of high-electrification systems facing persistent supply constraints worldwide.

\section{Acknowledgments}
This publication is based on research supported by the Templeton World Charity Foundation, Inc. 
(funder DOI 501100011730) under the grant
\url{https://doi.org/10.54224/32645}.
The survey protocol was reviewed and deemed exempt by the Institutional Review Board (IRB) of the American University of Beirut (IRB ID: SBS-2024-0538). We thank Raghid Farhat for her valuable contributions to the survey collection efforts.

\newpage
\bibliographystyle{unsrt}

\bibliography{bibliography}

\newpage
\appendix
\section{Representative profiles} \label{app:app_rep_prof}
\begin{figure}
    \centering
    \includegraphics[width=.9\linewidth]{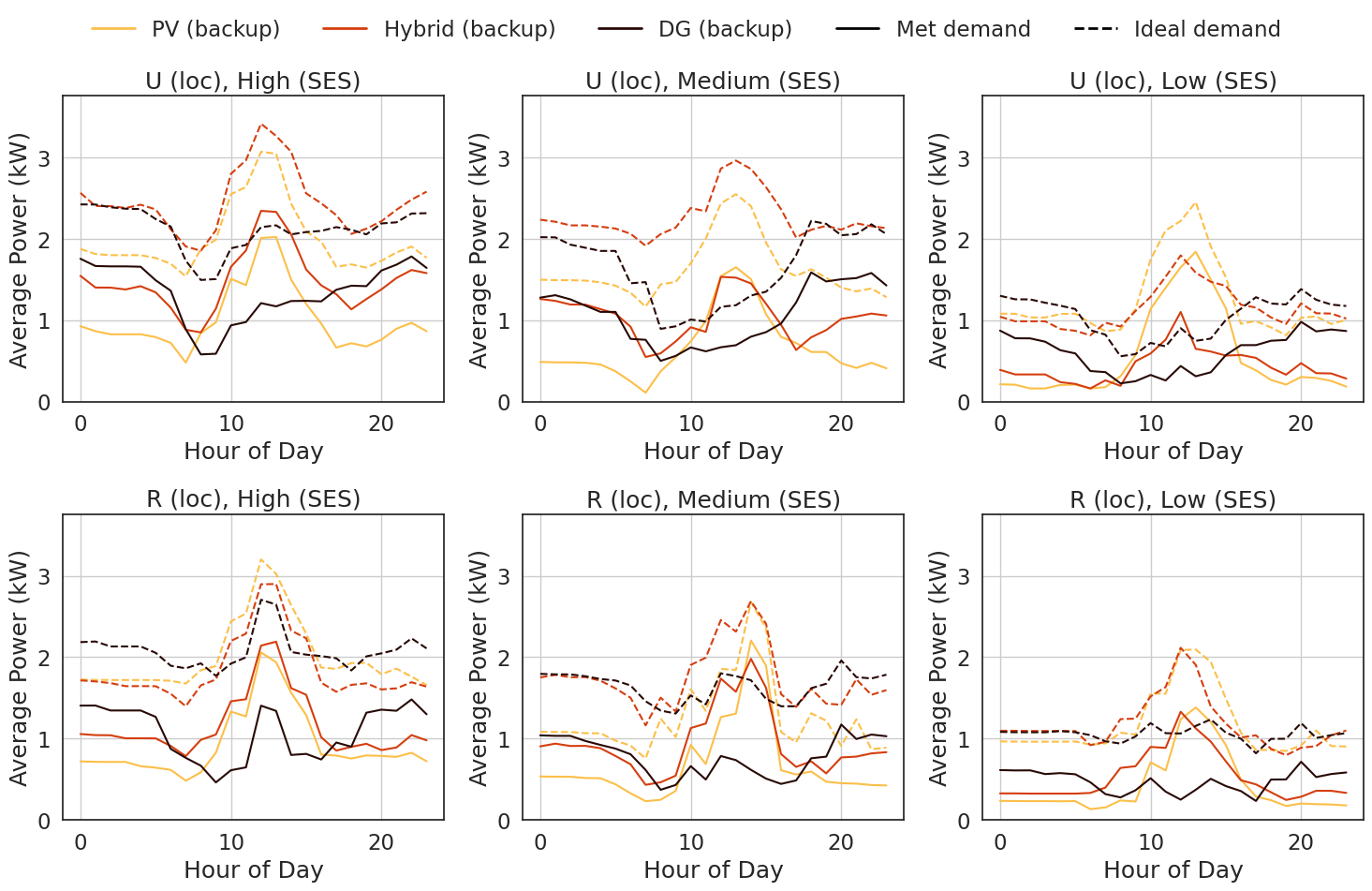}
    \caption{Daily load profiles in summer}
    \label{fig:summer_all}
\end{figure}
\begin{figure}
    \centering
    \includegraphics[width=.9\linewidth]{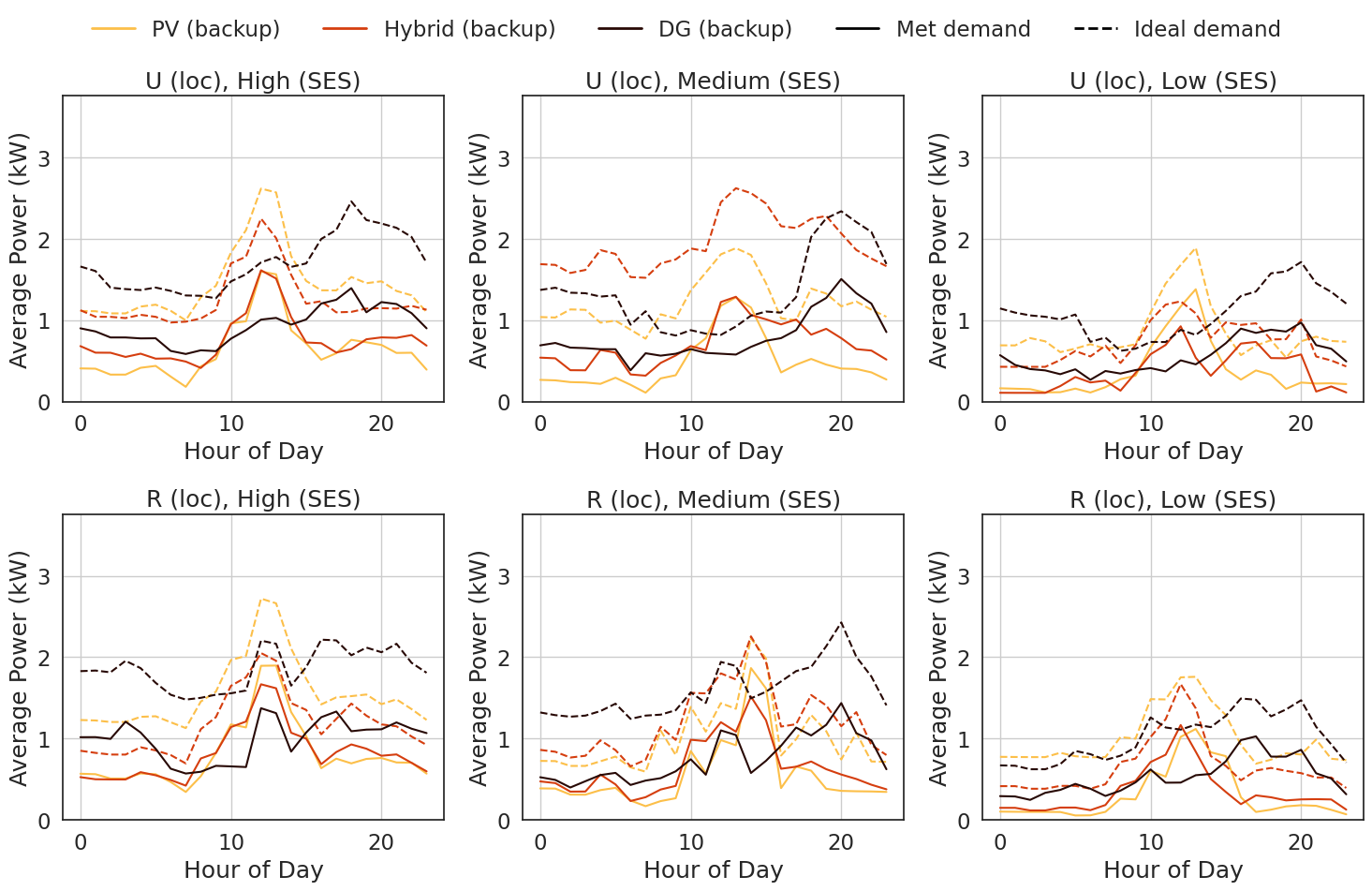}
    \caption{Daily load profiles in winter}
    \label{fig:winter_all}
\end{figure}
\begin{figure}
    \centering
    \includegraphics[width=.9\linewidth]{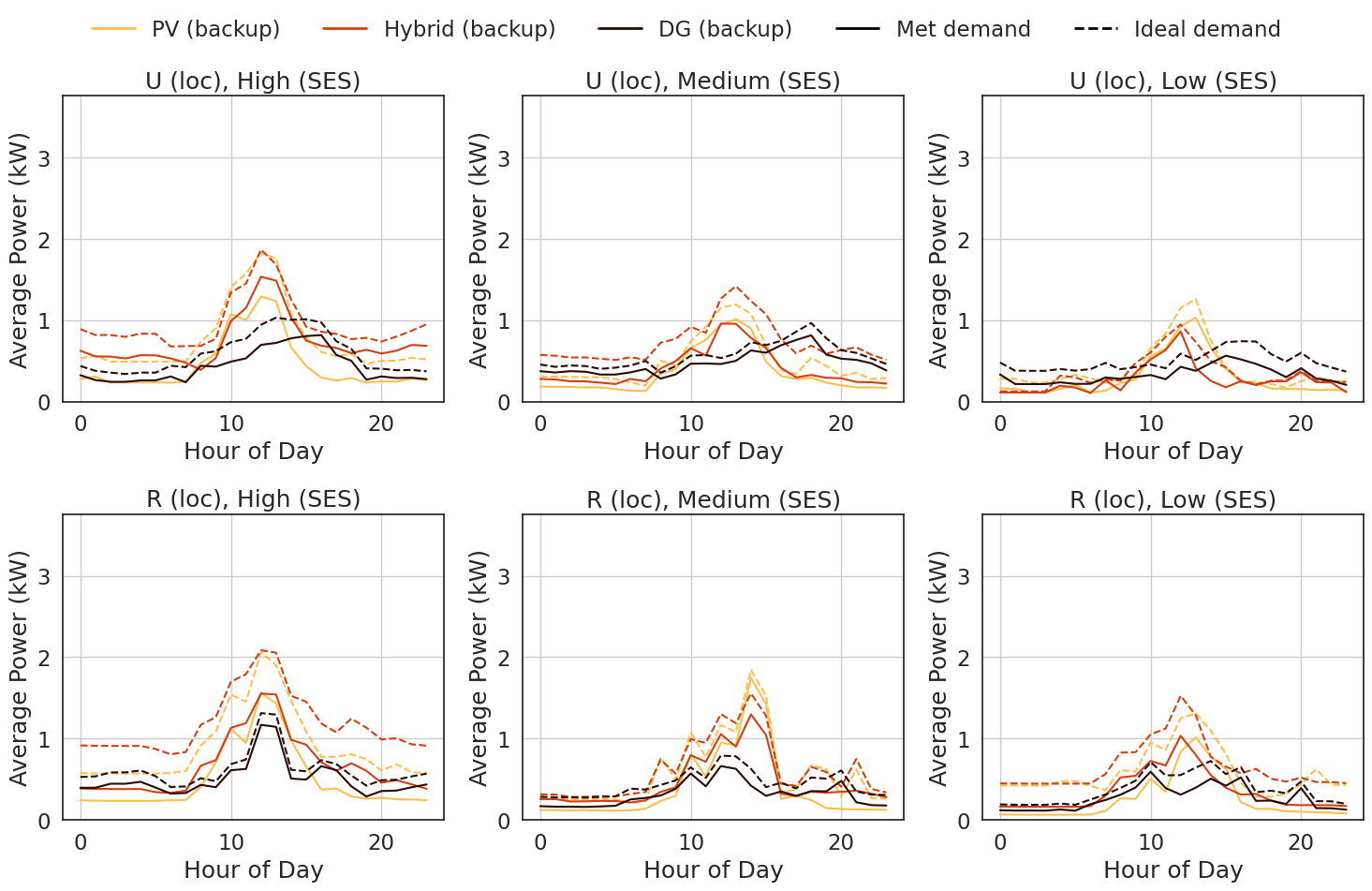}
    \caption{Daily load profiles in spring}
    \label{fig:spring_all}
\end{figure}
\begin{figure}
    \centering
    \includegraphics[width=.9\linewidth]{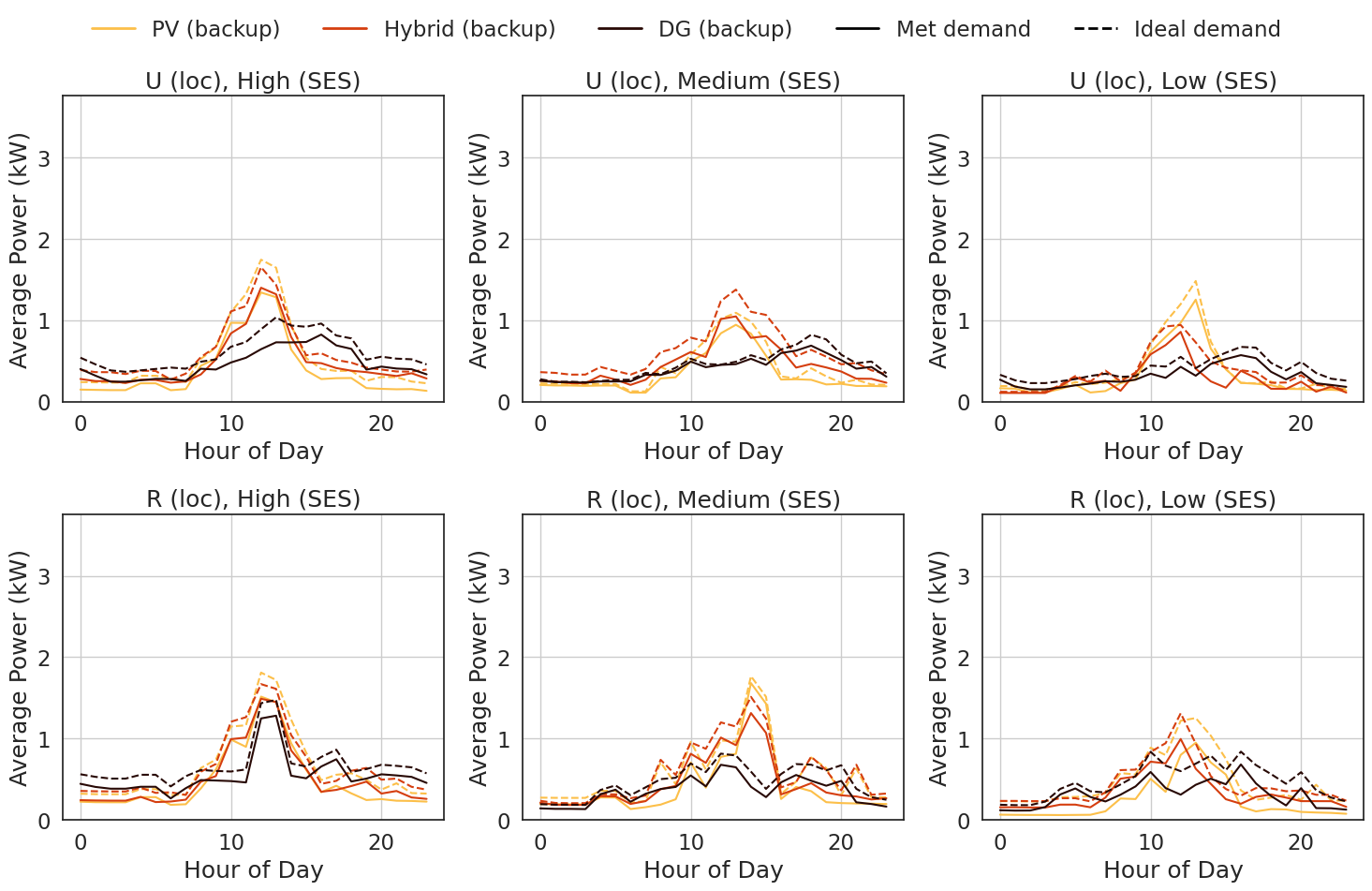}
    \caption{Daily load profiles in fall}
    \label{fig:fall_all}
\end{figure}
\FloatBarrier
\newpage

\section{Survey questions} \label{app:questions}
\begin{enumerate}
    \item General household information:
    \begin{itemize}
        \item How many persons live in the household?
        \item What is the area of the household in sqm?
        \item How many rooms are there in the household?
        \item Do you own a solar water heater? \textit{If not, why? Choose one or more:}
        \begin{itemize}
            \item Lack of funds
            \item Lack of information
            \item Already owns a non-solar water heater
            \item Lack of space
            \item Inadequate solar exposure
            \item Conflict with neighborhood agreements
        \item Doesn't want to
        \item Rents the unit
        \item Other, specify
        \end{itemize}
        \item What percentage of the lighting in your household is LED?
    \end{itemize}
    \item Backup generation:
    \begin{itemize}
        \item Indicate which of the below generators you have access to:
        \begin{itemize}
            \item Rooftop PV and battery
            \item Neighborhood diesel generator \textit{if yes, indicate monthly cost in USD and consumption in kWh}
            \item Personal diesel generator \textit{if yes, indicate monthly cost in USD and consumption in kWh}
            \item EDL
            \item Other, specify
        \end{itemize}
        \item If you do not have a household PV system, why? Choose one of the below:
        \begin{itemize}
            \item Lack of funds
            \item Lack of information
            \item Inconvenient change in deman
            \item Lack of space
            \item Inadequate solar exposure
            \item Conflict with neighborhood agreements
        \item Doesn't want to
        \item Rents the unit
        \item Other, specify
        \end{itemize}
        \item If you have a household PV-battery system, please specify:
        \begin{itemize}
            \item The number of panels or the capacity of your PV system
            \item The year of installation
            \item The number of batteries or the storage capacity of your battery system
            \item The year of installation or last change
            \item The type of batteries owned
        \end{itemize}
    \end{itemize}
    \item Demand:
    \begin{itemize}
        \item Fill the table (Table \ref{tab:demand_quest}) by specifying the number of appliances owned, and indicating by 1 the hour at which the appliance is turned on, and 0 when it is turned off 
    \end{itemize}
    \item Unmet demand:
    \begin{itemize}
        \item Fill the table (Table \ref{tab:demand_quest}) by specifying the number of appliances owned, and indicating by 1 the hour at which you would like to turn on the appliance for additional use, and 0 when the additional use ceases.
    \end{itemize}
\end{enumerate}

\begin{table}[]
\begin{tabular}{llllllll}
\hline
Timing & Appliances & \begin{tabular}[c]{@{}l@{}}Number\\  owned\end{tabular} & 0 & 1 & ... & 22 & 23 \\ \hline
Year & Refrigerator &  &  &  &  &  &  \\
Year & Freezer &  &  &  &  &  &  \\
Year & Clothes washer &  &  &  &  &  &  \\
Year & Clothes dryer &  &  &  &  &  &  \\
Year & Dishwasher &  &  &  &  &  &  \\
Year & Microwave &  &  &  &  &  &  \\
Year & Electric oven &  &  &  &  &  &  \\
Year & Electric cooktop &  &  &  &  &  &  \\
Year & Water pump &  &  &  &  &  &  \\
Year & Lighting &  &  &  &  &  &  \\
Year & Other 1 &  &  &  &  &  &  \\
Year & Other 2 &  &  &  &  &  &  \\ \hline
Season & Inverter AC &  &  &  &  &  &  \\
Season & Non-inverter AC &  &  &  &  &  &  \\
Season & Electric heater &  &  &  &  &  &  \\
Season & Irrigation pump &  &  &  &  &  &  \\
Season & Electric water heater &  &  &  &  &  &  \\
Season & Other 1 (seasonal) &  &  &  &  &  &  \\
Season & Other 2 (seasonal) &  &  &  &  &  &  \\ \hline
\end{tabular}
\caption{Table for demand questions}
\label{tab:demand_quest}
\end{table}
\newpage
\printcredits



\end{document}